# Removing leakage-induced correlated errors in superconducting quantum error correction

M. McEwen,[1,2] D. Kafri,[3] Z. Chen,[2] J. Atalaya,[3] K. J. Satzinger,[2] C. Quintana,[2] P. V. Klimov,[2] D. Sank,[2] C. Gidney,[2] A. G. Fowler,[2] F. Arute,[2] K. Arya,[2] B. Buckley,[2] B. Burkett,[2] N. Bushnell,[2] B. Chiaro,[2] R. Collins,[2] S. Demura,[2] A. Dunsworth,[2] C. Erickson,[2] B. Foxen,[2] M. Giustina,[2] T. Huang,[2] S. Hong,[2] E. Jeffrey,[2] S. Kim,[2] K. Kechedzhi,[3] F. Kostritsa,[2] P. Laptev,[2] A. Megrant,[2] X. Mi,[2] J. Mutus,[2] O. Naaman,[2] M. Neeley,[2] C. Neill,[2] M. Niu,[3] A. Paler,[4,5] N. Redd,[2] P. Roushan,[2] T. C. White,[2] J. Yao,[2] P. Yeh,[2] A. Zalcman,[2] Yu Chen,[2] V. N. Smelyanskiy,[3] John M. Martinis,[1] H. Neven,[2] J. Kelly,[2] A. N. Korotkov,[2,6] A. G. Petukhov,[2] and R. Barends[2]

[1] *Department of Physics, University of California, Santa Barbara, CA 93106, USA*
[2] *Google, Santa Barbara, CA 93117, USA*
[3] *Google, Venice, CA 90291, USA*
[4] *Johannes Kepler University, 4040 Linz, Austria*
[5] *University of Texas at Dallas, Richardson, TX 75080, USA*
[6] *Department of Electrical and Computer Engineering, University of California, Riverside, CA 92521, USA*

(Dated: September 29, 2020)

Quantum computing can become scalable through error correction, but logical error rates only decrease with system size when physical errors are sufficiently uncorrelated. During computation, unused high energy levels of the qubits can become excited, creating leakage states that are long-lived and mobile. Particularly for superconducting transmon qubits, this leakage opens a path to errors that are correlated in space and time. Here, we report a reset protocol that returns a qubit to the ground state from all relevant higher level states. We test its performance with the bit-flip stabilizer code, a simplified version of the surface code for quantum error correction. We investigate the accumulation and dynamics of leakage during error correction. Using this protocol, we find lower rates of logical errors and an improved scaling and stability of error suppression with increasing qubit number. This demonstration provides a key step on the path towards scalable quantum computing.

Quantum error correction stabilizes logical states by operating on arrays of physical qubits in superpositions of their computational basis states [1–3]. Superconducting transmon qubits are an appealing platform for the implementation of quantum error correction [4–13]. However, the fundamental operations, such as single-qubit gates [14, 15], entangling gates [16–20], and measurement [21] are known to populate non-computational levels, creating a demand for a reset protocol [22–27] that can remove leakage population from these higher states without adversely impacting performance in a large scale system. Directly quantifying leakage during normal operation presents another challenge, as optimizing measurement for detecting multiple levels is hard to combine with high speed and fidelity. This calls for analysis methods that use the errors detected during the stabilizer code's operation to find and visualize undesired correlated errors.

Here we introduce a multi-level reset gate using an adiabatic swap operation between the qubit and the readout resonator combined with a fast return. It requires only 250 ns to produce the ground state with a fidelity of over 99% for qubits starting in any of the first three excited states, with gate error accurately predicted by an intuitive semi-classical model. It is straightforward to calibrate and robust to drift due to the adiabaticity. Further, it uses only existing hardware as needed for normal operation and readout, and does not involve strong microwave drives that might induce crosstalk, making it attractive for large scale systems.

We benchmark the reset gate using the bit-flip error correction code [5] and measure growth and removal of leakage in-situ. By purposefully injecting leakage, we also quantify the gate's impact on errors detected in the code. Finally, we introduce a technique for computing the probabilities of error pairs, which allows identifying the distinctive patterns of correlations introduced by leakage. We find applying reset reduces the magnitude of correlations. We use these pair probabilities to inform the identification and correction of errors, improving the code's performance and stability over time.

The multi-level reset gate consists of the three distinct stages dubbed "swap", "hold", and "return" (Fig. 1a). First, we swap all qubit excitations to the resonator by adiabatically sweeping the qubit frequency to ∼ 1 GHz below the resonator frequency. We then hold the qubit below the resonator while excitations decay to the environment. Finally, we return the qubit diabatically to its initial frequency.

Pulse engineering of the "swap" stage is critical to achieving efficient population transfer. We adopt a fast quasi-adiabatic approach [28], where the qubit frequency changes rapidly when far detuned from the resonator level crossing but changes slowly when near the level crossing, see Supplementary Information. Since the frequency changes more slowly near the level crossing than a linear ramp, the probability of a diabatic error $P_D^{(s)}$ can be upper bounded by a Landau-Zener transition. This gives $P_D^{(s)} \ll \exp\left(-(2\pi g)^2 t_{\text{swap}}/\Delta f\right) \sim 10^{-3}$, where $t_{\text{swap}} = 30$ ns, $\Delta f = 2.5$ GHz is the total qubit frequency change and $g \approx 120$ MHz is the qubit-resonator coupling [29].

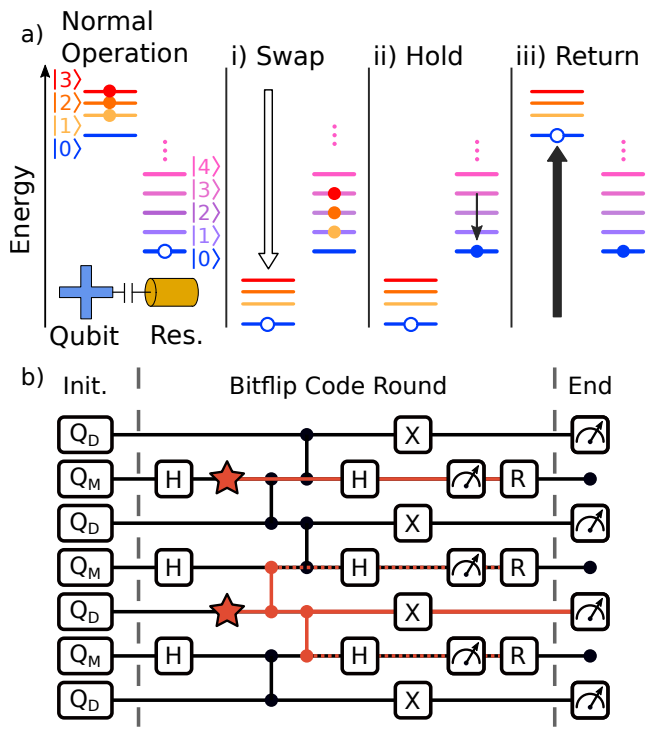

Figure 1. **Removing leakage with reset.** (a) Schematic of the multi-level reset protocol. The qubit starts with population in its first three excited states (closed circles), with the readout resonator in the ground state (open circle). (i) The qubit is swept adiabatically past the resonator to swap excitations. (ii) Resonator occupation decays to the environment while the qubit holds. (iii) After the resonator is sufficiently depleted, the qubit returns diabatically to its operating frequency. The total duration of the reset protocol is about 250 ns. (b) Circuit for the bit-flip stabilizer code including reset (R). Measure qubits ($Q_M$) cyclically apply parity measurements to neighbouring data qubits ($Q_D$) using Hadamard (H) and CZ gates. When introducing reset, leakage errors (stars) may be removed from both measure and data qubits, either directly or via transport through the CZ gates (red lines).

The "hold" stage of the protocol is primarily described by resonator photon decay. This decay follows $\exp(-\kappa t_{\text{hold}}) \sim 10^{-3}$, with $t_{\text{hold}} \sim 300$ ns and $\kappa \sim 1/(45\text{ns})$ the resonator decay rate. The qubit's excitation number remains mostly unchanged during the hold below the resonator as Purcell decay [30] through the resonator is small. For swap lengths below 30 ns the adiabaticity of the swap transition breaks down, and the system enters the "hold" stage in a superposition of the two adiabatic eigenstates. As a result, the probability undergoes coherent Rabi oscillations, which causes an incomplete reset and manifests itself as fringes.

If a single photon remains in the qubit-resonator system, the "return" stage of the protocol can be well described by a Landau-Zener transition. Achieving diabaticity is limited by the finite bandwidth of the control system. We can estimate an effective detuning velocity $\nu_r = \frac{1}{h}\frac{d}{dt}(E_{01} - E_{10}) = \Delta f / t_r$ using the typical ramp time-scale $t_r = 2$ ns. The probability of the desired diabatic transition is then $P_D^{(r)} = \exp[-(2\pi g)^2/\nu_r] \approx 0.6$. This description can be further extended to the multi-photon case using the Landau-Zener chain model [31].

Combining the semi-classical descriptions of each stage, we can identify two error channels in the reset of a single excitation. The first channel corresponds to the photon adiabatically swapping into the resonator, but then surviving over the hold time and adiabatically transitioning back to the qubit during the return. This is the dominant error channel, with probability $(1 - P_D^{(s)})e^{-\kappa t_{\text{hold}}}(1 - P_D^{(r)}) \sim 5 \cdot 10^{-4}$. The second channel corresponds to a failed initial swap of the qubit photon, followed by a diabatic transition during the return. The probability of this error is small, approximately $P_D^{(s)} P_D^{(r)} \ll 10^{-4}$. The reset dynamics of the 2- and 3-states is similar, with multiple adiabatic transitions moving 2 and 3 photons to the resonator respectively, after which they undergo rapid decay.

We experimentally test our reset gate on a Sycamore processor [29], consisting of an array of flux-tunable superconducting transmon qubits [4, 32] with tunable couplers [17, 29, 33, 34]. Each qubit is coupled to a readout resonator with strength $g \approx 120$ MHz, and having a frequency $\sim 1.5$ GHz below the qubit. Resonators are coupled to the outside through a Purcell filter [35].

The reset gate is implemented using flux-tuning pulses to steer the qubit's frequency to interact with the resonator, see Fig. 2a. The selected qubit has an idle frequency of 6.09 GHz and a nonlinearity of 200 MHz. The qubit starts at its idle frequency, moves past the resonator at 4.665 GHz, and is held 1 GHz below it, followed by a fast return to the idle frequency. We define the reset error as the likelihood of producing any state other than the ground state. The dependence of reset error on swap duration is shown in Fig. 2b for the cases when the qubit is initialized to $|1\rangle$, $|2\rangle$, and $|3\rangle$. We find that the reset error for all of the initialized states decreases until it reaches the readout visibility floor at about 30 ns swap length. This floor of 0.2% is measured independently as the ground state measurement error after after projecting the qubit into $|0\rangle$ using a prior measurement. We notice oscillations in the data which arise from an incomplete swap and are reproduced by the theoretical model results. In Fig. 2c, we keep the swap duration fixed at 30 ns and vary the hold duration. We find that the reset error decreases exponentially until it reaches the readout visibility floor, with a decay that is compatible with eter choice, stemming from the adiabaticity of the gate, highlights the protocol's robustness to drift and noise. This makes it amenable for use in large-scale systems. Finally, the demonstrated ability to simultaneously remove occupation from the 2- and 3-state makes this protocol a prime candidate for mitigating leakage in quantum error correction.

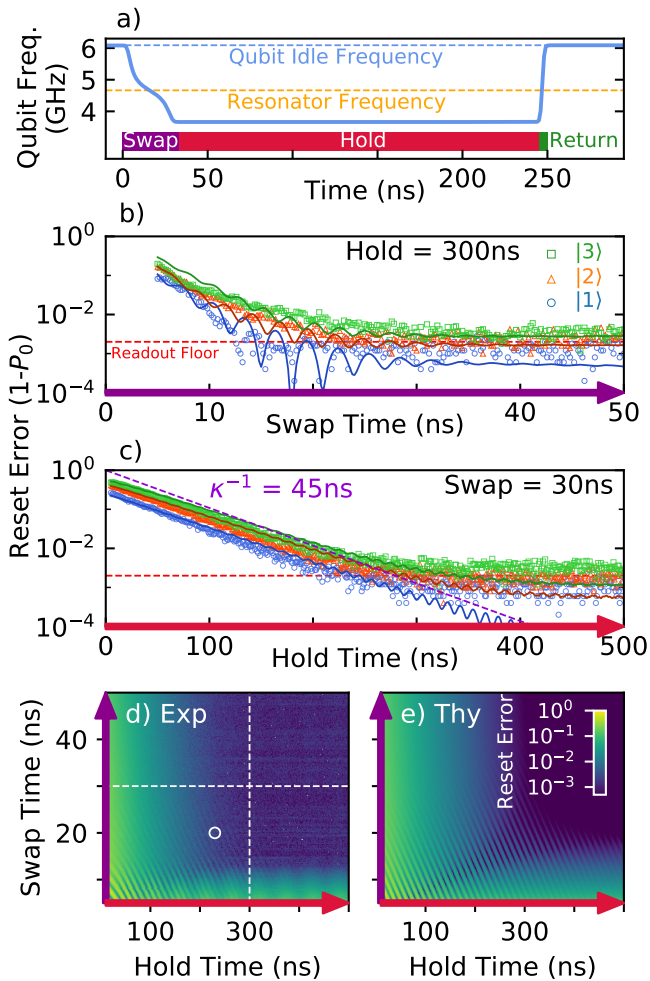

Figure 2. **Reset gate benchmarking.** (a) The qubit frequency trajectory for implementing reset consists of three stages. We plot the ground state infidelity when resetting the first three excited states of the qubit versus swap (b) and vs hold times (c). We include experimental data (points) and theory prediction (solid lines). Reset error versus swap and hold for experiment (d) and theory (e) show a wide range of optimal parameters. Dashed white lines indicate are linecuts for (b) and (c). White circle indicates the point of operation.

We now benchmark this protocol in the bit-flip stabilizer code [5], a precursor to the surface code. Here, a fast cycle of Hadamard, entangling, and measurement gates is repeated [Fig. 1b] to extract parity measurements to stabilize the logical state. We note the addition of X gates on the data qubits to depolarize energy relaxation error. Since the reset protocol is designed to unconditionally prepare the ground state, and thus remove all quantum data, we apply it only on the measure qubits immediately after readout.

We implement a 21 qubit chain on a Sycamore processor (inset of Fig. 3). We start by directly measuring the growth of leakage to $|2\rangle$ by running the code for a number of rounds and terminating with a measurement that can resolve $|2\rangle$ on all qubits. We average over 40 random initial states for the data qubits, and find that the population of $|2\rangle$ grows and saturates. In the absence of reset, the measure qubits build up a larger $|2\rangle$ state population than the data qubits, indicating that readout is a significant source of leakage during operation.

We fit a simple rate equation model and calculate the leakage ($\gamma_\uparrow$) and decay ($\gamma_\downarrow$) rates for the $|2\rangle$ state popula-

$1/\kappa = 45$ns. We show the landscape of the reset error for the qubit initialized in $|1\rangle$, experimentally in Fig. 2d, and the model results in Fig. 2e. For a wide choice of parameters above a minimum swap and hold duration, the ground state can be achieved with high fidelity: Experimentally we are limited by readout and theoretically the deviation from the ground state is below $10^{-3}$. We also note that the majority of error is favorably in the computational basis, which stabilizer codes can naturally identify and correct. The landscape involving other parameters can be found in the Supplementary Information.

The data and model results in Fig. 2 show that one can reset a qubit within 250 ns to the ground state with an error of around $10^{-3}$. Moreover, the insensitivity to param-tion [15]. Applying reset to the measure qubits breaks the established pattern of growth and requires a different fitting procedure, see Supplementary Information. We find a forty-fold increase in $\gamma_\downarrow$ for measure qubits with the addition of reset. We also find a 2.4x increase in $\gamma_\downarrow$ for data qubits, indicating transport of leakage population from data to measure qubits. We understand this effect as arising naturally in our CZ gate [33], which requires a condition that also places $|21\rangle$ and $|03\rangle$ on resonance, where the $|2\rangle$ is on the lower frequency qubit. Where a data qubit is below the measure qubit in frequency, transport of $|2\rangle$ from the data qubit to $|3\rangle$ in the measure qubit can occur, where it is subsequently removed by reset.

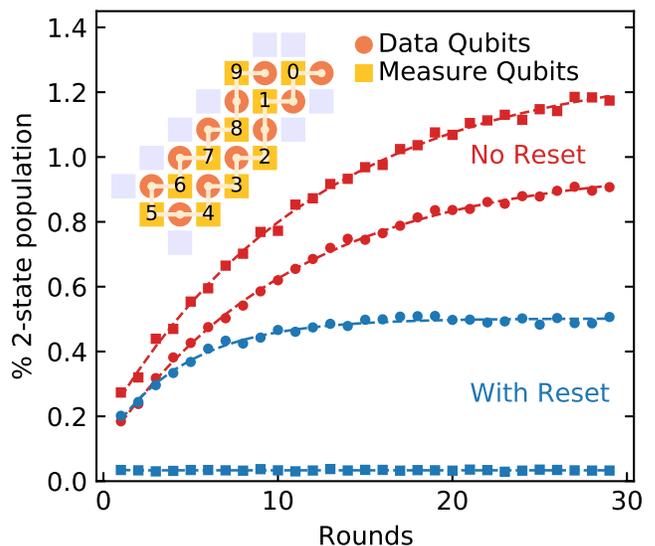

Figure 3. **Leakage during the Bit-flip code.** The growth in $|2\rangle$ population vs. stabilizer code length. The circuit is run for a number of rounds and terminated with a readout sensitive to $|2\rangle$ population. The experimental data is averaged over measure or data qubits and fitted to an exponential (dashed lines) to extract rates. The inset shows the 21 qubit chain as implemented on the Sycamore device.

To visualize the pattern of errors that leakage produces, we now inject $|2\rangle$ into the stabilizer code at specific locations. We insert a complete rotation between $|1\rangle$ and $|2\rangle$ on a single qubit immediately after the first Hadamard gates in round 10 of a 30 round experiment. As the data qubits are in either $|0\rangle$ or $|1\rangle$, and measure qubits are in an equal superposition of $|0\rangle$ and $|1\rangle$ after the Hadamard, the amount of injected $|2\rangle$ is the same for both measure and data qubits on average. Fig. 4 shows the fraction of error detection events, which represents the portion of runs where a given stabilizer measurement reports an unexpected result, indicating an error occurred [5]. Injected leakage produces two distinct effects; a pair of detection events at injection, and a tail of correlated detection events over the lifetime of the leakage state. As with discrete bit-flip errors, the initial pairs of detection events appear sequentially in time for injection on measure qubits, while for data qubits both adjacent measure qubits report error (gray arrows).

The detection event fractions for all qubits are shown in the insets and cross-sections are shown in the main figure. We note that the value of the detection event fraction deviates for the first round due to initialization, and for the last round as data qubit measurements are involved [5]. As can be seen in Fig. 4a, the insertion of leakage in measure qubit 5 (see inset of Fig. 3 for its location) creates two adjacent peaks at a detection event fraction of 0.5, as the injection produces a random readout result in round 10. This is followed by a clear tail of anomalously high levels of detection events that slowly decays over many rounds, indicating errors that are correlated in time. When applying reset, the errors on all measure qubits are more uniform, and the increase in detection events for the first nine rounds becomes flattened. Importantly, the slow decay in errors is no longer visible as the detection event fraction drops to the baseline immediately after the initial pair of detection events. We also insert leakage in the data qubit between measure qubits 4 and 5, see Fig. 4b. We again notice an increase of detection events that slowly decays, now on both neighbouring measure qubits. The error decreases more rapidly with reset, corroborating our prior observation that higher level states can migrate to measure qubits. In addition, we notice a small increase in detection events around the leakage injection in qubits 3 and 6 in the case of no reset, further indicating that higher level states can move between qubits. We notice for both cases a small odd-even oscillation in the data, which we understand as arising from the fact that the injected $|1\rangle$ to $|2\rangle$ rotation does not affect the data qubit when it is in state $|0\rangle$. Since the X gates on data qubits swap $|0\rangle$ and $|1\rangle$ in each round, we see a higher likelihood of bit error from energy relaxation in odd rounds after the injection.

The data in Fig. 4 show that the reset protocol can remove large populations of leakage in measure qubits and helps to decrease leakage in data qubits, thereby strongly suppressing time-correlated tails of detection events. This result also raises the question how higher level state occupations that naturally arise during the stabilizer codes lead to correlated errors.

To further quantify this, we analyze the correlations between detection events that arise during normal code operation using the error graph [5], see Fig. 5a. We model detection events as arising from independent random processes that flip pairs of measurements [In preparation, Google AI Quantum and Collaborators]. The probability $p_{ij}$ of the process that flips measurements $i$ and $j$ can be obtained from the observed correlations between detection events,

$$p_{ij} = \frac{1}{2} - \frac{1}{2}\sqrt{1 - \frac{4\left(\langle x_i x_j \rangle - \langle x_i \rangle \langle x_j \rangle\right)}{1 - 2\langle x_i \rangle - 2\langle x_j \rangle + 4\langle x_i x_j \rangle}}, \quad (1)$$

where $x_i = 1$ if there is a detection event at a given measurement $i$ and $x_i = 0$ otherwise. Here $\langle x \rangle$ denotes averaging $x$ over many experimental realizations.

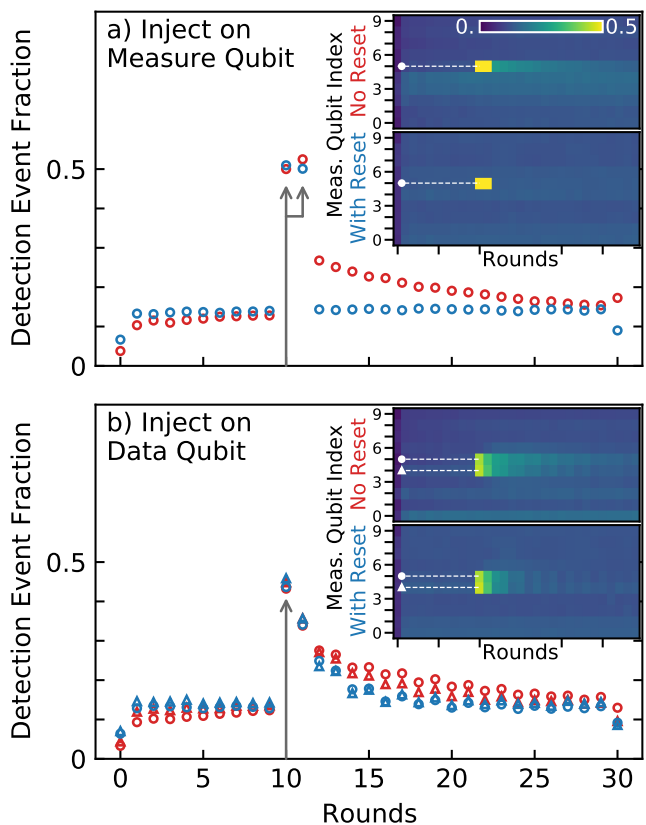

Figure 4. **Injection of leakage.** Detection event fraction when a full $|1\rangle \to |2\rangle$ rotation is inserted during round 10 after the first Hadamards (a) on measure qubit 5 and (b) on the data qubit between measure qubits 4 (circles) and 5 (triangles) during round 10. Insets show the event fraction across all measure qubits, indicating the traces plotted in the main figure (dashed lines). See Fig. 3 inset for qubit locations.

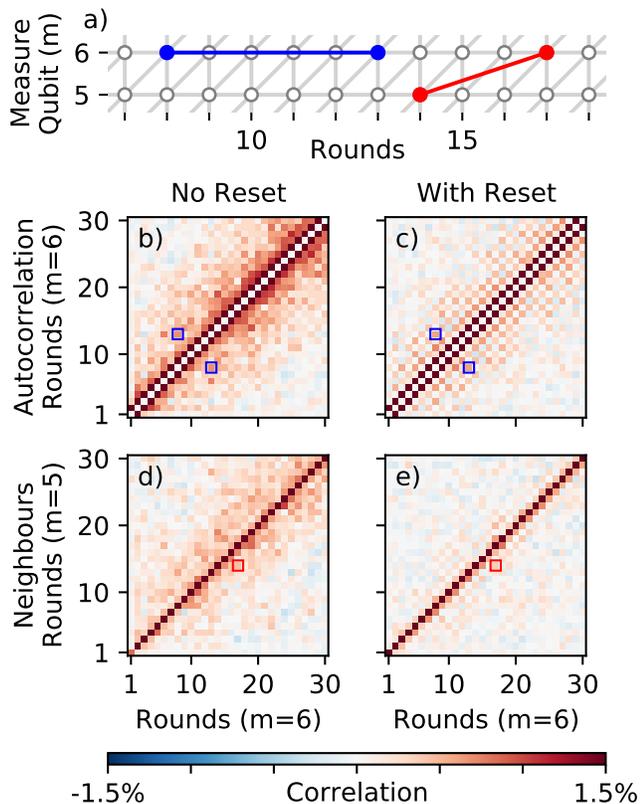
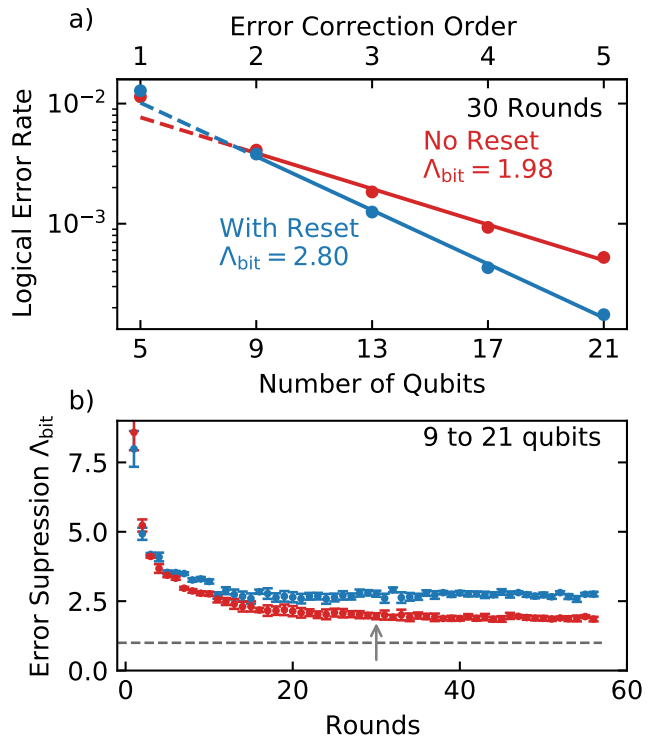

Figure 5. **Correlations caused by leakage.** $p_{ij}$ matrices show the strength of non-local correlations in the detected errors and their reduction with reset. (a) The error graph for the bit-flip code, highlighting examples of non-local correlations on both space and time, indicating their corresponding $p_{ij}$ elements below (boxes). (b,c) The matrix for time-correlations on measure qubit 6, with and without reset. (d,e) The matrix for cross correlations between measure qubits 5 and 6, with and without reset.

Figure 6. **Logical code performance.** (a) The logical error rate for 30 rounds vs system size. The error suppression factor $\Lambda_{\rm bit}$ is fitted to the data from nine qubits up. (b) $\Lambda_{\rm bit}$ versus code depth, showing that with reset logical error suppression is improved consistently. The threshold for the bit-flip code (unity) is shown as a dashed line. The arrow indicates the data in (a).

In Fig. 5, we visualize $p_{ij}$ and show autocorrelations for measure qubit 6 and cross correlations between measure qubits 5 and 6. The standard error correction model assumes that detection events occur only in local pairs. For detection events occurring on the same measure qubit, we expect only correlations between adjacent rounds, corresponding to elements adjacent to the main diagonal ($p_{i,i\pm1}$) of Fig. 5b,c. For detection events occurring on neighboring measure qubits, we expect only correlations between the qubits in the same round, or adjacent rounds due to the staggered placement of CZ gates. This corresponds to non-zero elements only on and immediately below the main diagonal of Fig. 5d,e. In contrast, without reset, we find that significant unexpected correlations appear (left panels), covering distances of over 10 rounds. With reset, these long-range correlations are mostly removed (right panels). This reveals an underlying checkerboard pattern which arises similarly to the aforementioned odd-even oscillations (see Supplementary Information).

Having shown the reset protocol removes leakage and suppresses long-distance correlations, we now look at logical error rates. We run the stabilizer code to a given number of rounds, and feed the detection events into a minimum weight perfect matching algorithm [36] that identifies and keeps track of errors to return the corrected logical state.

We perform the experiment on 21 qubits, and use sub-sampling [5][In preparation, Google AI Quantum and Collaborators] to evaluate performance for smaller subsets of the code to extract scaling performance. We average over these subsets to calculate an average probability of logical error for 5, 9, 13, 17 qubits at each number of rounds. We use the $p_{ij}$ elements to set the weights for the matching algorithm. We convert the probability of a logical error ($P_L$) at a given number of rounds $k$ to a logical error rate $\epsilon = [1 - (1 - 2P_L)^{(1/k)}]/2$ [37] for the number of rounds $k$, shown at 30 rounds in Fig. 6a. Here, the logical error rate is plotted from 5 to 21 qubits, corresponding to an error correction order of $n = 1$ to 5, meaning at least $n+1$ errors must occur to cause a logical error. The error rate of the bit-flip code in the absence of correlations should be exponentially suppressed with $\epsilon \propto 1/\Lambda_{\text{bit}}^{n+1}$.

We find that the logical error rate decreases with number of qubits, with an exponential dependence from 9 qubits up. We exclude data at 5 qubits, due to the degraded performance which we attribute to boundary effects impacting matching performance [38].

We plot the $\Lambda_{\text{bit}}$ versus rounds in Fig. 6b. A constant logical error rate should produce a $\Lambda_{\text{bit}}$ that is independent of round number. In practice, effects including the buildup of leakage, the thermalization of data qubits, and time boundary effects will produce an higher apparent $\Lambda_{\text{bit}}$ prior to saturation. We observe $\Lambda_{\text{bit}}$ decaying over 30 rounds toward a saturation value of 1.98 for the case of no reset. With reset, $\Lambda_{\text{bit}}$ stabilizes faster, within 10 rounds, to a higher value of 2.80. Notably, error suppression is enhanced despite the time added to the cycle by reset, where data qubits are exposed to additional decoherence. This highlights the importance of removing the time-correlated errors induced by leakage, as seen in Fig 5.

We observe the logical performance stabilizing to values of $\Lambda_{\text{bit}} > 1$, and that the addition of reset improves both the stable performance and rate with which the code approaches stability. Moreover, we see deviations from ideal behaviour where experiments are small in number of qubits or rounds, where boundary effects may artificially inflate $\Lambda_{\text{bit}}$. This highlights that error suppression is a property that asymptotically emerges with space and time.

In summary, we introduce a reset protocol that uses existing hardware to remove higher level states and test it using the bit-flip stabilizer code. We show that reset mitigates leakage-induced long-time correlated errors and significantly improves logical error suppression. While optimizing gates and readout to have minimal leakage is a necessary strategy, the correlated nature of the error that leakage induces makes reset protocols critical for quantum error correction.

# Supplementary information for "Removing leakage-induced correlated errors in superconducting quantum error correction"

M. McEwen,[1,2] D. Kafri,[3] Z. Chen,[2] J. Atalaya,[3] K. J. Satzinger,[2] C. Quintana,[2] P. V. Klimov,[2] D. Sank,[2] C. Gidney,[2] A. G. Fowler,[2] F. Arute,[2] K. Arya,[2] B. Buckley,[2] B. Burkett,[2] N. Bushnell,[2] B. Chiaro,[2] R. Collins,[2] S. Demura,[2] A. Dunsworth,[2] C. Erickson,[2] B. Foxen,[2] M. Giustina,[2] T. Huang,[2] S. Hong,[2] E. Jeffrey,[2] S. Kim,[2] K. Kechedzhi,[3] F. Kostritsa,[2] P. Laptev,[2] A. Megrant,[2] X. Mi,[2] J. Mutus,[2] O. Naaman,[2] M. Neeley,[2] C. Neill,[2] M. Niu,[3] A. Paler,[4,5] N. Redd,[2] P. Roushan,[2] T. C. White,[2] J. Yao,[2] P. Yeh,[2] A. Zalcman,[2] Yu Chen,[2] V. N. Smelyanskiy,[3] John M. Martinis,[1] H. Neven,[2] J. Kelly,[2] A. N. Korotkov,[2,6] A. G. Petukhov,[2] and R. Barends[2]

[1]*Department of Physics, University of California, Santa Barbara, CA 93106, USA*
[2]*Google, Santa Barbara, CA 93117, USA*
[3]*Google, Venice, CA 90291, USA*
[4]*Johannes Kepler University, 4040 Linz, Austria*
[5]*University of Texas at Dallas, Richardson, TX 75080, USA*
[6]*Department of Electrical and Computer Engineering, University of California, Riverside, CA 92521, USA*

(Dated: September 18, 2020)

## I. RESET GATE PARAMETERS

The multi-level reset gate has five main parameters that determine its shape, shown in Fig. 2a in the main text: the swap, hold, and return durations, and two parameters that determine the swap trajectory. We calibrate the swap and hold times as described in the main text, and we use a minimum return time imposed by filtering of 2 ns so as to maximize $P_D^{(r)}$.

The adiabatic swap we use follows the quasi-adiabatic approach of Ref. [1] with a modification explained below. In this approach the pulse shape $f_q(t)$ is designed in terms of the "control angle" $\theta$ on the Bloch sphere of states $|01\rangle$ and $|10\rangle$, $\tan(\theta) = 2g/(f_q - f_r)$, $0 < \theta < \pi$, where $f_q(t)$ is the qubit frequency, $f_r$ is the resonator frequency, and $g$ is the qubit-resonator coupling. However, the reset gate has to operate not only for the initial state $|1\rangle$ of the qubit, but also for the states $|2\rangle$ and $|3\rangle$ having stronger couplings. Moreover, there are three relevant resonance conditions for these cases: $f_q = f_r$, $f_q = f_r + \eta$, and $f_q = f_r + 2\eta$, where $\eta \simeq 200$ MHz is the qubit nonlinearity. We therefore use a phenomenological approach and design the pulse $f_q(t)$ as in Ref. [1], but for somewhat different coupling and resonator frequency. We replace $g$ and $f_r$ with free parameters $\mu$ and $f_\text{swap}$, respectively, and design a pulse $f_q(t)$ and optimize experimental performance of the reset gate over $\mu$ and $f_\text{swap}$.

For clarity, we now describe the process outlined in Ref. [1] in more detail. The pulse shape $f_q(t)$ is parametrized as $d\tilde{\theta}/dx = (\tilde{\theta}_\text{fin} - \tilde{\theta}_\text{in}) \sum_{n=1}^{3} \lambda_n [1 - \cos(2\pi n x)]$, where $\sum_{n=1}^{3} \lambda_n = 1$, $\tan(\tilde{\theta}) = 2\mu/(f_q - f_\text{swap})$, and $0 < \tilde{\theta} < \pi$. Here, the initial and final values of $\tilde{\theta}$ are defined as $\tan(\tilde{\theta}_\text{in}) = 2\mu/(f_\text{idle} - f_\text{swap})$ and $\tan(\tilde{\theta}_\text{fin}) = 2\mu/(f_\text{hold} - f_\text{swap})$. We also define a dimensionless "natural" time $x$, $0 \leq x \leq 1$, for which the Rabi frequency is constant. This dimensionless time is related to the physical time $t$ as $t = t_\text{swap} \int_0^x \sin\tilde{\theta}(x')\, dx' / [\int_0^1 \sin\tilde{\theta}(x')\, dx']$. We first calculate $\tilde{\theta}(x)$ analytically, then calculate $t(x)$ numerically, and then use numerical interpolation to find $\tilde{\theta}(x(t))$. Finally, the qubit trajectory is obtained as $f_q(t) = f_\text{swap} + 2\mu \cot[\tilde{\theta}(t)]$. We use $\lambda_1 = 1.15$, $\lambda_2 = -0.2$, and $\lambda_3 = 0.05$, similar to the values used in Ref. [2]. We set $f_\text{hold}$ to be 1 GHz below the readout resonator frequency $f_r$ to minimize the hybridization of levels.

Figure S1 shows the error of the reset gate as a function of the parameter $f_\text{swap}$ and the swap duration $t_\text{swap}$ for the qubit initial states (a) $|1\rangle$, (b) $|2\rangle$, and (c) $|3\rangle$. For the initial state $|1\rangle$ and small values of swap duration, we see that performance is optimized near $f_\text{swap} = f_r$, whereas at higher values of $t_\text{swap}$ the dependence is obscured by the readout floor. As expected, for the initial states $|2\rangle$ and $|3\rangle$, the optimal value of $f_\text{swap}$ is significantly higher than $f_r$. Nevertheless, we set $f_\text{swap} = f_r$; this gives an acceptable performance for all initial states at sufficiently long $t_\text{swap}$.

The parameter $\mu$ affects the slope of the pulse shape $f_q(t)$ at $f_\text{swap}$, and therefore the adiabaticity. A bigger value of $\mu$ (compared with $g$) increases the slope near $f_\text{swap}$ but decreases the slope at the "sides" of the pulse, thus broadening the frequency range around $f_\text{swap}$ over which the slope is approximately constant. This results in a bigger diabatic error for the initial state $|1\rangle$ (for which $\mu = g$ is supposed to be the optimum), but decreases the error for the initial states $|2\rangle$ and $|3\rangle$, for which the first resonance occurs at $f_r + \eta$ and/or $f_r + 2\eta$ (the effective physical couplings in this case are also higher). Figure S2 shows the error landscape as a function of $\mu$ and $t_\text{swap}$ for the three initial states. As expected, for the initial states $|2\rangle$ and $|3\rangle$, having values of $\mu$ larger than $g = 120$ MHz is preferred. As a compromise, we choose $\mu = 150$ MHz. This value not only relaxes the frequency selectivity for the crossing point, but also decreases sensitivity to noise or drift in the frequency bias.

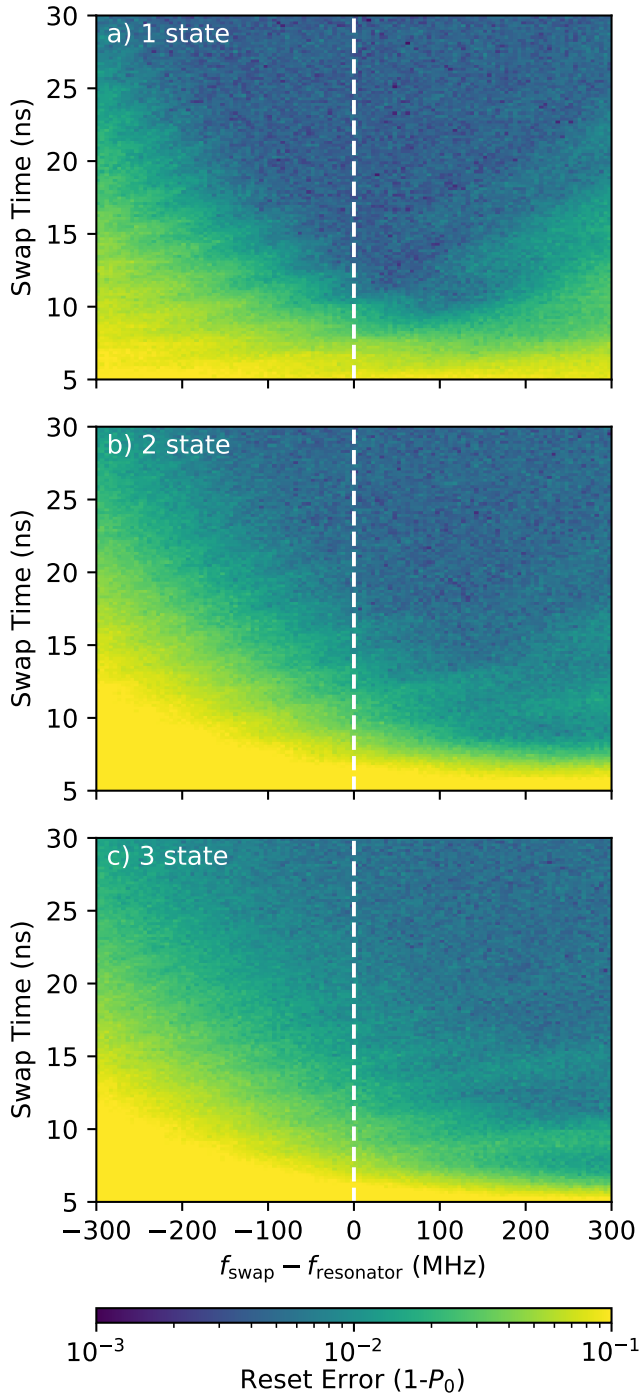

Figure S1. **Dependence on swap frequency.** Reset performance when applied to a qubit initialized in $|1\rangle$ (a), $|2\rangle$ (b) and $|3\rangle$ (c) versus $f_{\rm swap} - f_{\rm resonator}$. For short swap lengths and on input $|1\rangle$, performance is approximately optimal for $f_{\rm swap} = f_{\rm resonator}$ (dashed line). Reset of higher states also involve transitions when the qubit is above the readout resonator due to the negative nonlinearity, producing distinct landscapes. At long swap lengths, this dependence is obscured by the readout floor, but degraded performance on $|2\rangle$ and $|3\rangle$ states is visible for $f_{\rm swap} < f_{\rm resonator}$.

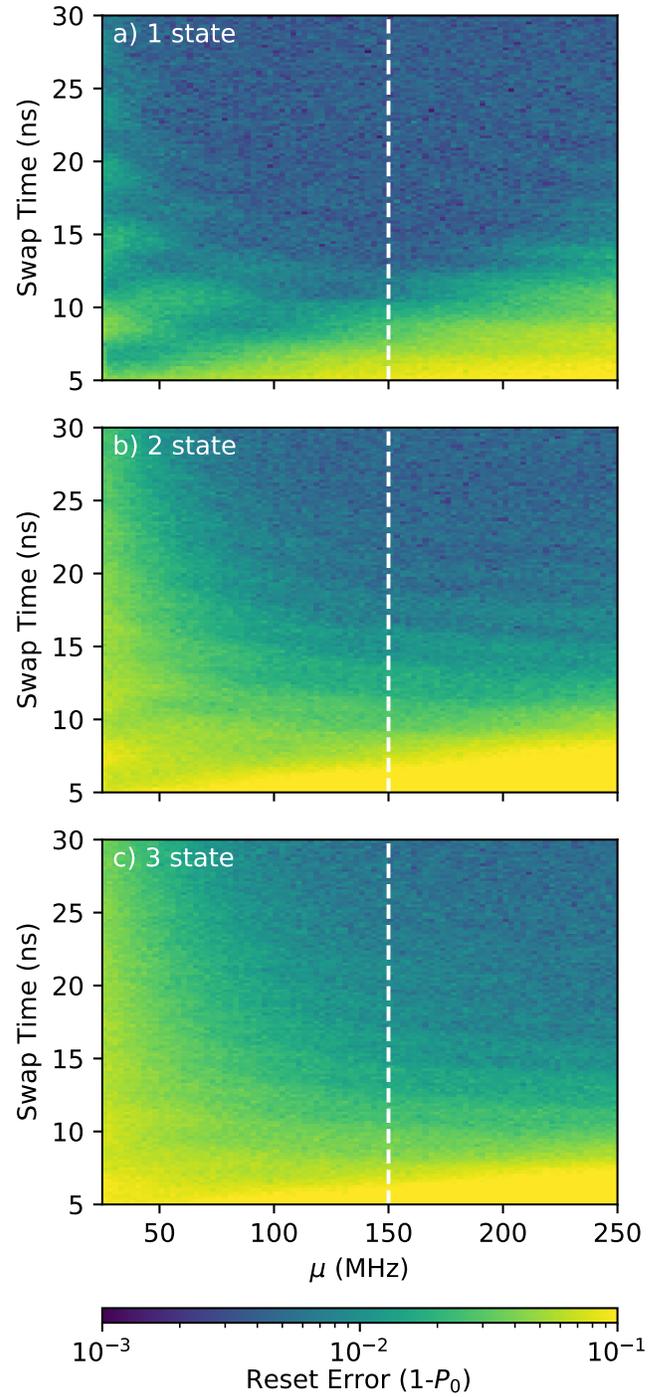

Figure S2. **Dependence on adiabatic slope parameter.** Reset performance when applied to qubits in states $|1\rangle$ (a), $|2\rangle$ (b) and $|3\rangle$ (c) versus the adiabatic slope parameter $\mu$. For short swap lengths, performance by different input states show different profiles depending on the number of transitions involved. We choose a value of $\mu$=150 MHz as a compromise between these three cases (dashed lines). At long swap lengths, this dependence is mostly obscured by the readout floor, but degraded performance on $|2\rangle$ and $|3\rangle$ states is visible at smaller values of $\mu$.

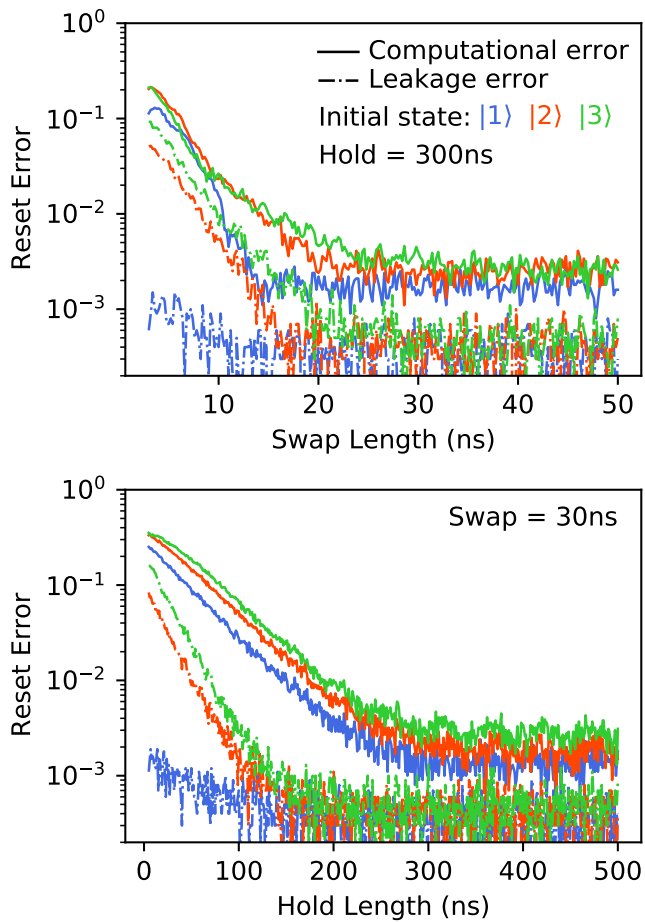

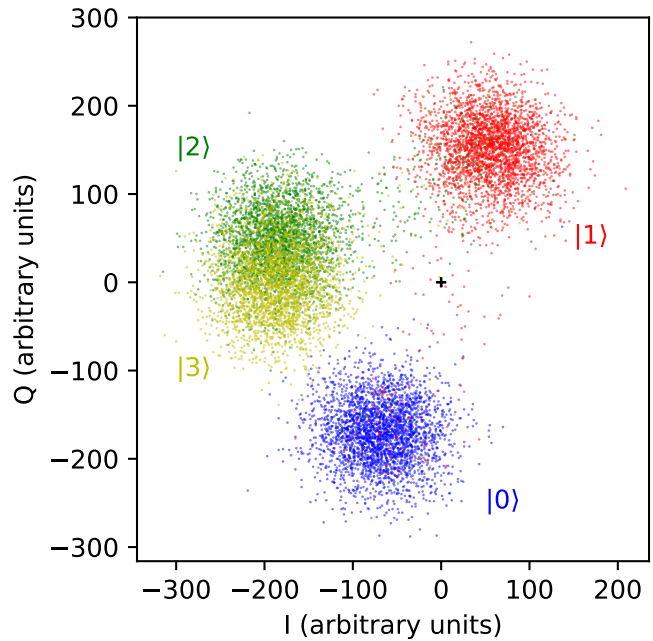

Figure S4. **Readout for distinguishing leakage.** Demodulated in-phase (I) and quadrature (Q) components measured using a readout optimized for distinguishing computational states and leakage states. The qubit is prepared in either $|0\rangle$, $|1\rangle$, $|2\rangle$ or $|3\rangle$. This illustrates our ability to distinguish the two computational states from each other and from higher level states with high fidelity.

Figure S3. **Computational vs. Leakage error.** Reset error separated into computational error ($P_1$, solid lines) and leakage error ($P_2 + P_3$, dash-dotted lines) for the reset gate applied to the first three excited states. We see that computational error accounts for the majority of error in all cases.

## II. LEAKAGE ERROR AND SUPPRESSION

We can distinguish between two kinds of error produced by the reset gate. We define the 'computational error' as the probability that the qubit is in the $|1\rangle$ state after reset, and 'leakage error' as the probability that it remains in a higher state ($|2\rangle$ and $|3\rangle$). In the context of error correction, computational error is preferred as the code naturally identifies and corrects for errors within the computational basis, and the single and two qubit gates involved are calibrated primarily for computational basis states.

In Fig. S3, we show reset performance separated into computational and leakage error. For short swap and hold lengths, we can see that the ratio of leakage to computational error depends on the initial state, with higher states producing more leakage error as expected. We also see a higher rate of reduction for leakage error than for computational error with hold time, reflecting the higher rate of energy relaxation from higher states. At swap and hold lengths long enough to reach the readout floor, we see that the leakage error is around 10x lower than the computational error for all states, indicating that the dominant error source is computational error.

As in Fig. 3 in the main article, we can distinguish these two error types using a readout optimized for detecting higher states. A representative example of such a readout result is shown in Fig. S4. The qubit is repeatedly prepared in $|0\rangle$, $|1\rangle$, $|2\rangle$ or $|3\rangle$, and the complex readout signal is measured and demodulated. Each shot is plotted as a single point, colored by the prepared state, allowing us to evaluate the readout fidelity for various states and to calibrate the discrimination of states. This readout was optimized to distinguish the two computational states from leakage states with high fidelity, but does not attempt to distinguish the leakage states $|2\rangle$ and $|3\rangle$ from each other.

For such a readout, we find the readout floor in Fig. 2 in the main article by heralding; performing two sequential measurements on the qubit, postselecting on $|0\rangle$ on the first measurement and calculating the fidelity of measuring $|0\rangle$ again on the second measurement.

## III. RATE EQUATIONS FOR LEAKAGE DURING CODE OPERATION

In Fig. 3 of the main text, we measure the growth of leakage with rounds $k$ during the bit-flip code using a readout similar to that shown in Fig. S4. We fit the leakage population to an exponential to extract parameters for a rate equation [3],

$$P_{|2\rangle}(k) = p_\infty \left(1 - e^{-\Gamma k}\right) + p_0 e^{-\Gamma k} \qquad \text{(S1)}$$

$$\Gamma = \gamma_\uparrow + \gamma_\downarrow \qquad p_\infty = \frac{\gamma_\uparrow}{\Gamma} \qquad \text{(S2)}$$

The rates are displayed in Table S1, showing an increase in effective leakage decay rate when reset is applied. As seen in Fig. 3 in the main text, applying reset to the measure qubits breaks the established behaviour for growth of leakage. In order to estimate $\gamma_\downarrow$, we therefore assume a value of $\gamma_\uparrow$ equal to the case of no reset, and a value of $p_\infty$ given by the average error for measure qubits across all rounds.

Table S1. Effective leakage growth and decay rate per stabilizer round, using Eq. S1. For the case of measure qubits with reset, $\gamma_\downarrow$ was estimated from the value of $p_\infty$ when assuming $\gamma_\uparrow$ equal to the case of no reset (asterisk).

|            |         | $\gamma_\uparrow$ | $\gamma_\downarrow$ | $p_\infty$ |
|------------|---------|-------------------|---------------------|------------|
| No Reset   | Data    | 0.09%             | 9.1%                | 0.97%      |
|            | Measure | 0.11%             | 8.1%                | 1.30%      |
| With Reset | Data    | 0.11%             | 22.1%               | 0.50%      |
|            | Measure | 0.11%*            | 328%*               | 0.03%*     |

## IV. THE CHECKERBOARD PATTERN IN THE $p_{ij}$-MATRIX

There is a clear checkerboard pattern visible in Fig. 5c in the main article, in which the values of the correlation matrix $p_{ij}$ for measure qubit 6 are larger for correlations spanning an odd number of rounds. For edges spanning an even number of rounds, the correlations are smaller and can even be negative. A similar but less pronounced checkerboard pattern can be seen in Fig. 5e for the cross-correlation between measure qubits 5 and 6. In fact, (b) and (c) display similar patterns, but these are visually masked by the presence of significant leakage-induced correlations. Both edges shown in Fig. 5a span odd numbers of rounds and show values larger than neighbouring values in the $p_{ij}$ matrix.

This checkerboard pattern is caused by correlations between energy relaxations on the same data qubit. The mechanism of the correlation is illustrated in Fig. S5. An energy relaxation event on a data qubit, $|1\rangle \to |0\rangle$, produces a pair of detection events on the neighbouring measure qubits (red circles in Fig. S5). Subsequent $X$-gates applied to the data qubit each round (see Fig. 1b in the main article) alternate the qubit state: $|0\rangle \to |1\rangle \to$ $|0\rangle \to |1\rangle \to \ldots$ . As a result, the qubit can relax again 1, 3, 5, ... rounds later, while relaxation after 2, 4, 6, ... rounds is unlikely. This creates a positive correlation between the errors separated by an odd number of rounds and negative correlation for separation by an even number of rounds, producing the checkerboard pattern. The correlations gradually decay with increasing separation. The checkerboard pattern is more pronounced in Fig. 5c as a single measure qubit is affected by both neighbouring data qubits, while in Fig. 5e the checkerboard pattern is caused by only the one data qubit between the measure qubits 5 and 6.

## V. STATISTICS AND POSTSELECTION IN THE BIT-FLIP CODE

When benchmarking performance in the bit-flip code, we average over a large number of realizations, including over randomly chosen initial states for the data qubits. For the leakage populations shown in Fig. 3 of the main text, we chose 20 random initial bitstrings for the data qubits, and repeated the experiment 5000 times for each bitstring. For data shown in Figs. 4, 5 and 6, we chose 40 random bitstrings and repeated the experiment 1000 times for each bitstring for 40 000 total realizations. However, the probabilities of logical errors are smaller at low numbers of rounds, requiring additional averaging to reduce statistical error. For runs with 10 or fewer rounds, we therefore chose 100 random initial bitstrings and took 10 000 repetitions at each bitstring, for 1 000 000 total realizations.

Over these large numbers of runs, we see a small number of short 'events' where the detection fractions are significantly elevated compared to the average. We are investigating these effects [In preparation, Google AI Quantum and Collaborators][In preparation, McEwen et al.]. These events are not representative of the normal functioning of the device and so we postselect them out using the following procedure. We calculate the logical error for each time-ordered realization. We then calculate a moving average of the logical error over 30 realisations. The total average logical error is below 3%, but during events the moving average can reach 50%. We choose a threshold of 25% to identify the start of an event, and remove around 1000 realisations for each event identified. This removes approximately 0.8% of the data.

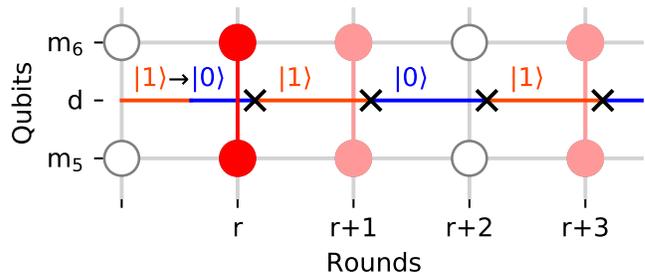

Figure S5. **Odd-even periodicity of energy relaxation events.** The state of a data qubit $d$ is flipped each round by an X gate (black). An energy relaxation event $|1\rangle \to |0\rangle$ in $d$ in round $r$ produces detection events (red circles) in the neighboring measure qubits $m_5$ and $m_6$. As the next energy relaxation event can occur only when $d$ is in $|1\rangle$, future energy relaxation errors will be preferentially separated by an odd number of rounds from the initial event (pink circles), producing an alternating pattern of correlations.